\documentclass[aps,prd,preprint,showpacs,superscriptaddress]{revtex4-1}
\usepackage{amsmath}
\usepackage{latexsym}
\usepackage{amsfonts}
\usepackage{graphicx}

\begin{document}
\title{Thermodynamics of scalar--tensor theory with non-minimally derivative coupling}

\author{Yumei Huang}
\email{huangymei@gmail.com}
\affiliation{Department of Astronomy, Beijing Normal University, Beijing 100875, China}

\author{Yungui Gong}
\email{yggong@mail.hust.edu.cn}
\affiliation{MOE Key Laboratory of Fundamental Quantities Measurement, School of Physics, Huazhong University of Science and Technology,
Wuhan, Hubei 430074, China}

\author{Dicong Liang}
\email{904186306@qq.com}
\affiliation{School of Physics, Huazhong University of Science and Technology,
Wuhan, Hubei 430074, China}

\author{Zhu Yi}
\email{yizhuhust@sina.cn}
\affiliation{School of Physics, Huazhong University of Science and Technology,
Wuhan, Hubei 430074, China}

\begin{abstract}
With the usual definitions for the entropy and the temperature
associated with the apparent horizon,
we show that the unified first law on the apparent horizon  is equivalent to the Friedmann equation for the scalar--tensor
theory with non-minimally derivative coupling. The second law of thermodynamics on the apparent horizon
is also satisfied. The results support a deep and fundamental connection between gravitation, thermodynamics, and quantum theory.
\end{abstract}

\preprint{1504.01271}

\maketitle

\section{Introduction}

The discovery of black hole thermodynamics \cite{Bardeen:1973gs} has shown a deep connection
between gravitation and thermodynamics.
In particular, the black hole temperature, which is proportional to the surface gravity
at the event horizon, and Hawking radiation \cite{Hawking:1974sw} tell us that this relation may
be linked to quantum gravity \cite{Wald:1999vt}.
Instead of proportional to the volume, the Bekenstein--Hawking entropy is equal to
one quarter of the area of the event horizon
of the black hole measured in Planck units \cite{Bekenstein:1973ur,Hawking:1974sw}.
Based on this area law of entropy, Bekenstein
then argued a universal entropy bound for a weakly self-gravitating physical system in
an asymptotically flat space-time \cite{Bekenstein:1980jp}. This led to the proposal of the
holographic principle \cite{'tHooft:1993gx,Susskind:1994vu,Witten:1998qj}. The holographic principle
was supported by the AdS/CFT correspondence, which states that the type IIB superstring theory
on AdS$_5\times S^5$ is equivalent to the $N = 4$ super-Yang--Mills theory with gauge group
$U(N)$ in four dimensions \cite{Maldacena:1997re}. The AdS/CFT correspondence relates a gravitational
theory in $d$-dimensional anti-de Sitter space with a conformal field theory living in a
$(d-1)$-dimensional boundary space. The Hawking radiation and the holography show that
the thermodynamic property of gravitation is unique. These special properties may provide
some physical insights into the nature of quantum gravity.
By applying the area law of entropy for all local acceleration horizons, it was found that
Einstein equation could be derived from the first law of thermodynamics \cite{Jacobson:1995ab}.
The relation was then discussed in cosmology, and
the equivalence between the first law of thermodynamics
and Friedmann equation was derived \cite{Cai:2005ra,Akbar:2006kj,Cai:2006rs,Gong:2006ma,Gong:2007md}.
The relation between thermodynamics and gravitation was discussed extensively in the literature, and the relation
holds also in more general theories of gravity
\cite{Jacobson:1995ab,Hayward:1997jp,Padmanabhan:2002sha,Padmanabhan:2003gd,Cai:2005ra,Akbar:2006kj,Cai:2006rs,
Gong:2006ma,Gong:2007md,Wang:2005pk,Eling:2006aw,Kothawala:2007em,Zhou:2007pz,Gong:2006sn,
Cai:2008gw,Hayward:2008jq,Padmanabhan:2009vy,Sharif:2012zzd,Sharif:2013tha,Padmanabhan:2013lpa,Binetruy:2014ela,Mitra:2015nba,Helou:2015yqa}.

The simplest generalization of Einstein's general relativity is Brans--Dicke theroy \cite{Brans:1961sx}.
In Brans--Dicke theory, gravitation is propagated by massless spin zero scalar field in addition to the massless spin 2 graviton.
The scalar degree of freedom can also arise upon compactification of higher dimensions.
In general, the scalar field $\phi$ is coupled to the curvature scalar $R$ as $f(\phi)R$.
More general couplings for the scalar field are also possible \cite{Horndeski:1974wa,Amendola:1993uh,Capozziello:1999uwa,Capozziello:1999xt}.
In Horndeski theory, the derivatives
of both the metric $g_{\mu\nu}$ and the scalar field $\phi$ are at most second order, and
the second derivative $\phi_{;\mu\nu}$ couples to the Einstein tensor by the
general form $f(\phi,X)G^{\mu\nu}\phi_{;\mu\nu}$, where $X=g^{\mu\nu}\phi_{,\mu}\phi_{,\nu}$ \cite{Horndeski:1974wa}.
However, the field equations are still second order in Horndeski theory. We can also consider the non-minimally
derivative coupling $\phi_{,\mu}\phi^{,\mu}R$, $\phi_{,\mu}\phi_{,\nu}R^{\mu\nu}$,
$\phi\Box\phi R$, $\phi\phi_{;\mu\nu} R^{\mu\nu}$, $\phi\phi_{,\mu}R^{;\mu}$ and $\phi^2\Box R$.
If we choose the non-minimally derivative coupling as $G^{\mu\nu}\phi_{,\mu}\phi_{,\nu}$,
then the field equations contain no more than second derivatives \cite{Sushkov:2009hk},
and the theory avoids the Boulware--Deser ghost \cite{Boulware:1972zf}.
With this choice of non-minimally derivative coupling, it was shown that the Higgs field
produced a successful slow-roll inflation without violating the unitarity bound and fine-tuning
the coupling constant $\lambda$ \cite{Germani:2010gm}.
The scalar--tensor theory with the non-minimally
derivative coupling $\omega^2 G^{\mu\nu}\phi_{,\mu}\phi_{,\nu}$ was discussed by lots of researchers
recently \cite{Daniel:2007kk,Saridakis:2010mf,Sushkov:2012za,Skugoreva:2013ooa,Germani:2010ux,
Germani:2011ua,DeFelice:2011uc,Tsujikawa:2012mk,Sadjadi:2010bz,Sadjadi:2013uza,Minamitsuji:2013ura,Granda:2009fh,
Granda:2010hb,Granda:2011eh,Jinno:2013fka,Sami:2012uh,Anabalon:2013oea,Rinaldi:2012vy,Koutsoumbas:2013boa,
Cisterna:2014nua,Huang:2014awa,Bravo-Gaete:2013dca,Bravo-Gaete:2014haa,Bruneton:2012zk,Feng:2013pba,Feng:2014tka,Dalianis:2014sqa,Dalianis:2014nwa,Dalianis:2015aba,Yang:2015pga}.

In this paper, we discuss the thermodynamics of the scalar--tensor theory with non-minimally derivative coupling
$\omega^2 G^{\mu\nu}\phi_{,\mu}\phi_{,\nu}$. The paper is organized as follows. In section II, we
review the scalar--tensor theory with non-minimally derivative coupling. The relation between the first
law of thermodynamics and Friedmann equation is presented in section III. We discuss the second law
of thermodynamics in section IV and conclusions are drawn in Section V.

\section{The scalar--tensor theory with non-minimally derivative coupling}

The action for the general scalar--tensor theory with non-minimally derivative coupling
which contains only up to second derivatives is
\begin{equation}
\label{action2}
S=\int {\rm d}^4x\sqrt{-g}\left[\frac{M_{pl}^2}{2}R-\frac{1}{2}(g^{\mu\nu}-\omega^2G^{\mu\nu})\partial_{\mu}\phi\partial_{\nu}\phi-V(\phi)\right]+S_b,
\end{equation}
where the Planck mass $M_{pl}^2=(8\pi G)^{-1}=\kappa^{-2}$, $w$ is the coupling constant with the dimension of inverse mass,
$V(\phi)$ corresponds to the scalar field potential, and
$S_b$ is the action for the background matter, which includes dust and radiation.
Varying the action (\ref{action2}) with respect to the metric $g_{\mu\nu}$,
we get
\begin{equation}
\label{eq2}
G_{\mu\nu}=R_{\mu\nu}-\frac{1}{2}g_{\mu\nu}R=\kappa^2 (T_{\mu\nu}^b+T_{\mu\nu}^c),
\end{equation}
where $T_{\mu\nu}^b$ is the energy-momentum tensor for the background matter,
and the effective energy-momentum tensor $T_{\mu\nu}^c$ for the scalar field is
\begin{equation}
\label{scaltmunu}
\begin{split}
T_{\mu\nu}^c=&\phi_{,\mu}\phi_{,\nu}-\frac{1}{2} g_{\mu\nu}(\phi_{,\alpha})^2-g_{\mu\nu}V(\phi)\\
&-\omega^2\left\{-\frac{1}{2}\phi_{,\mu}\phi_{,\nu}\,R+2\phi_{,\alpha}\nabla_{(\mu}\phi R^\alpha_{\nu)}+\phi^{,\alpha}\phi^{,\beta}R_{\mu\alpha\nu\beta}\right.\\
&+\nabla_\mu\nabla^\alpha\phi\nabla_\nu\nabla_\alpha\phi-\nabla_\mu\nabla_\nu\phi\Box\phi-\frac{1}{2}(\phi_{,\alpha})^2 G_{\mu\nu}\\
&\left. +g_{\mu\nu}\left[-\frac{1}{2}\nabla^\alpha\nabla^\beta\phi\nabla_\alpha\nabla_\beta\phi+\frac{1}{2}(\Box\phi)^2
-\phi_{,\alpha}\phi_{,\beta}\,R^{\alpha\beta}\right]\right\},
\end{split}
\end{equation}
Using the homogeneous and isotropic Friedmann--Robertson--Walker (FRW) metric,
\begin{equation}
\label{rw}
{\rm d}s^2=-{\rm d}t^2+\frac{a(t)^2}{1-kr^2}{\rm d}r^2+a(t)^2r^2({\rm d}\theta^2+\sin^2\theta {\rm d}\phi^2),
\end{equation}
where $k=0$, $-1$, $+1$ represents a flat, open, and closed universe respectively,
we obtain $T_{00}^c$ and $T_{11}^c$ as:
\begin{gather}
T_{00}^c=\frac{1}{2}\dot\phi^2+V(\phi)+\frac{9}{2}{\omega^2}H^2{\dot\phi}^2+\frac{3}{2}\omega^2\frac{k}{a^2}{\dot\phi}^2,\\
T_{11}^c=\frac{a^2}{1-k r^2}\left[\frac{1}{2}\dot\phi^2-V(\phi)-\frac{\omega^2}{2}{\dot\phi}^2\left(2\dot H+3H^2-\frac{k}{a^2}+\frac{4H\ddot\phi}{\dot\phi}\right)\right].
\end{gather}
Therefore, the effective energy density and pressure for the scalar field are given by:
\begin{gather}
\label{rhop1}
\rho_c=\frac{\dot\phi^2}{2}\left(1+{9}{\omega^2}H^2+3\omega^2\frac{k}{a^2}\right)+V(\phi),\\
\label{rhop2}
p_c=\frac{1}{2}\dot\phi^2-V(\phi)-\frac{\omega^2}{2}{\dot\phi}^2\left(2\dot H+3H^2-\frac{k}{a^2}+\frac{4H\ddot\phi}{\dot\phi}\right).
\end{gather}
The Friedman equations are
\begin{equation}
\label{frds21}
H^2+\frac{k}{a^2}=\frac{8\pi G }{3}(\rho_b+\rho_c)=\frac{8\pi G }{3}\left[\frac{1}{2}\dot\phi^2+V(\phi)+\frac{9}{2}{\omega^2}H^2{\dot\phi}^2
+\frac{3}{2}\omega^2\frac{k}{a^2}{\dot\phi}^2+\rho_b\right],
\end{equation}
\begin{equation}
\label{frds22}
\dot H-\frac{k}{a^2}=-4\pi G \left[\dot\phi^2+3\omega^2H^2\dot\phi^2
+2\omega^2\frac{k}{a^2}\dot\phi^2-\omega^2\frac{d}{dt}(H\dot\phi^2)+\rho_b+p_b\right].
\end{equation}
If the non-minimally derivative coupling is absent, $\omega^2=0$, we recover the standard result of Einstein gravity
with canonically scalar field.

\section{The relation between the first law of thermodynamics and Friedmann equation}

In this section, we discuss the equivalence between the first law of thermodynamics on the apparent horizon and Friedmann equation.
For a spherically symmetric space-time with the metric ${\rm d}s^2=g_{ab}{\rm d}x^a{\rm d}x^b+\tilde{r}^2{\rm d}\Omega^2$, where the unit spherical
metric ${\rm d}\Omega^2={\rm d}\theta^2+\sin^2\theta {\rm d}\phi^2$, the apparent horizon is defined as $f=g^{ab}\tilde{r}_{,a}\tilde{r}_{,b}=0$,
the dynamical surface gravity at the apparent horizon is $\kappa=\nabla_a\nabla^a\, \tilde r/2$ \cite{Hayward:1997jp},
and the Hawking temperature associated with the apparent horizon is $T=|\kappa|/2\pi$.
For the FRW metric (\ref{rw}), the apparent horizon is:
\begin{equation}
\label{ra}
\tilde r_A=(H^2+k/{a^2})^{-1/2}.
\end{equation}
The surface gravity at apparent horizon is
\begin{equation}
\kappa=\frac{1}{2}\nabla_a\nabla^a\,\tilde r=-\frac{1}{\tilde r_A}(1-\frac{\dot{\tilde r}_A}{2H\tilde r_A}),
\end{equation}
and the associated temperature is $T_A=\kappa/2\pi$. The entropy enclosed by the apparent horizon is $S_A=\pi\tilde r_A^2/G$.
Therefore, we have
\begin{equation}
\label{tds}
T_A{\rm d}S_A=-\frac{1}{2\pi\tilde r_A}\left(1-\frac{\dot{\tilde r}_A}{2H\tilde r_A}\right)\frac{2\pi{\tilde r}_A}{G}{\dot{\tilde r}_A}{\rm d}t=-\frac{1}{G}\left(1-\frac{\dot{\tilde r}_A}{2H\tilde r_A}\right){\dot{\tilde r}_A}{\rm d}t.
\end{equation}

For the scalar--tensor theory with non-minimally derivative coupling, the effective total energy density is
$\rho_{\rm {tot}}=\rho_{b}+\rho_{c}$. The total energy of the system inside the apparent horizon is $E=\rho_{\rm {tot}}V$,
where the volume $V=4\pi \tilde r_A^3/3$. So the energy change is
\begin{equation}
\label{derhot}
{\rm d}E=\rho_{\rm {tot}}{\rm d}V+V{\rm d}\rho_{\rm {tot}}=\rho_{\rm {tot}} 4\pi {\tilde r}_A^2 {\rm d}\tilde r_A+\frac{4}{3}\pi {\tilde r}_A^3{\rm d}\rho_{\rm {tot}},
\end{equation}
where ${\rm d}\tilde r_A=\dot{\tilde r}_A {\rm d}t$, and ${\rm d}\rho_{\rm {tot}}=\dot\rho_{\rm {tot}} {\rm d}t$.
By using the energy conservation for the total energy,
\begin{equation}
\label{idf}
\dot\rho_{\rm {tot}}+3H(\rho_{\rm {tot}}+p_{\rm {tot}})=0,
\end{equation}
we have
\begin{equation}
{\rm d}\rho_{\rm {tot}}=-3H(\rho_{\rm {tot}}+p_{\rm {tot}}){\rm d}t.
\end{equation}
Substituting the above result into Eq. (\ref{derhot}), we get
\begin{equation}
\label{eq34}
{\rm d}E=4\pi {\tilde r}_A^2\rho_{\rm {tot}}\dot{\tilde r}_A {\rm d}t-4\pi {\tilde r}_A^3H(\rho_{\rm {tot}}+p_{\rm {tot}}){\rm d}t.
\end{equation}
The work term $W{\rm d}V$ with $W=(\rho_{\rm {tot}}-p_{\rm {tot}})/2$ is
\begin{equation}
\label{wdv}
W{\rm d}V=2\pi{\tilde r}_A^2(\rho_{\rm {tot}}-p_{\rm {tot}}){\dot{\tilde r}_A}{\rm d}t.
\end{equation}
Applying the unified first law,
\begin{equation}
\label{det}
{\rm d}E=T_A{\rm d}S_A+W{\rm d}V,
\end{equation}
we get
\begin{equation}
\label{flo1}
4\pi {\tilde r}_A^3H(\rho_{\rm {tot}}+p_{\rm {tot}})\left(1-\frac{\dot{\tilde r}_A}{2H\tilde r_A}\right)
=\frac{1}{G}\left(1-\frac{\dot{\tilde r}_A}{2H\tilde r_A}\right){\dot{\tilde r}_A}.
\end{equation}
Taking the time derivative of the apparent horizon $\tilde r_A$ defined in Eq. (\ref{ra}), we get
\begin{equation}
\label{dotra}
\dot{\tilde r}_A=-{\tilde r_A}^3H\left(\dot H-\frac{k}{a^2}\right).
\end{equation}
Combining Eqs. (\ref{flo1}) and (\ref{dotra}), we get
\begin{equation}
\label{frd1}
\begin{split}
\dot H-\frac{k}{a^2}&=-{4\pi G}(\rho_{\rm {tot}}+p_{\rm {tot}}) \\
&=-{4\pi G}\left[{\dot\phi^2}+3\omega^2H^2\dot\phi^2+2\omega^2\frac{k}{a^2}\dot\phi^2-\omega^2\frac{d}{dt}(H\dot\phi^2)+\rho_b+p_b\right].
\end{split}
\end{equation}
Using the energy conservation equation (\ref{idf}), and integrating Eq. (\ref{frd1}), we obtain the Friedman equation
\begin{equation}
\label{frd2}
H^2+\frac{k}{a^2}=\frac{8\pi G }{3}\left[\frac{\dot\phi^2}{2}\left(1+{9}{\omega^2}H^2+3\omega^2\frac{k}{a^2}\right)+V(\phi)+\rho_b\right].
\end{equation}

Thus, we derive the Friedmann equation from the unified first law on the apparent horizon for the scalar--tensor theory
with non-minimally derivative coupling.

Now we would like to derive the unified first law starting from the Friedmann equation.
Substituting Eq. (\ref{frds22}) into Eq. (\ref{dotra}), we obtain
\begin{equation}
\label{dotrara3}
{\rm d}{\tilde r}_A={4\pi G}H\tilde r_A^3\left[{\dot\phi^2}+3\omega^2H^2\dot\phi^2+2\omega^2\frac{k}{a^2}\dot\phi^2-\omega^2\frac{d}{dt}(H\dot\phi^2)+\rho_b+p_b\right]{\rm d}t.
\end{equation}
We multiply $-[1-\dot{\tilde r}_A/(2H\tilde r_A)]/G $ both sides of Eq. (\ref{dotrara3}), then Eq. (\ref{dotrara3}) becomes
\begin{equation}
\label{tdsp}
\begin{split}
T_A{\rm d}S_A&=-\frac{1}{2\pi\tilde r_A}\left(1-\frac{\dot{\tilde r}_A}{2H\tilde r_A}\right)\cdot
{\rm d} \left(\frac{4\pi {\tilde r}_A^2}{4G}\right) \\
&=-{4\pi}H\left[{\dot\phi^2}+3\omega^2H^2\dot\phi^2+2\omega^2\frac{k}{a^2}\dot\phi^2-\omega^2\frac{d}{dt}(H\dot\phi^2)+\rho_b+p_b\right]{\tilde r}_A^3\left(1-\frac{\dot{\tilde r}_A}{2H\tilde r_A}\right){\rm d}t \\
&=-{4\pi}H(\rho_{\rm {tot}}+p_{\rm {tot}}){\tilde r}_A^3\left(1-\frac{\dot{\tilde r}_A}{2H\tilde r_A}\right){\rm d}t.
\end{split}
\end{equation}
Combining Eq. (\ref{tdsp}) with Eq. (\ref{wdv}), we get
\begin{equation}
\label{tfl}
\begin{split}
T_A{\rm d}S_A+W{\rm d}V&=-{4\pi}{\tilde r}_A^3H(\rho_{\rm {tot}}+p_{\rm {tot}}){\rm d}t+{4\pi}{\tilde r}_A^2\rho_{\rm {tot}}{\dot{\tilde r}_A}{\rm d}t\\
&=\frac{4\pi}{3}{\tilde r}_A^3 {\rm d}\rho_{\rm {tot}}+{4\pi}\rho_{\rm {tot}}{\tilde r}_A^2{\rm d} {\tilde r}_A={\rm d}E.
\end{split}
\end{equation}
So the unified first law is derived from the Friedmann equation together with the energy
conservation equation. Thus,
with the usual definitions for the entropy and the Hawking temperature
associated with the apparent horizon,
we show that the unified first law on the apparent horizon  is equivalent to the Friedmann equation for the scalar--tensor
theory with non-minimally derivative coupling.

\section{The second law of thermodynamics on the apparent horizon}

As discussed in the previous section, the entropy of the apparent horizon $S_A=A/(4G)=\pi\tilde r_A^2/G$, so
\begin{equation}
\dot S_A=\frac{2\pi\tilde r_A}{G}\dot{\tilde r}_A=-\frac{2\pi{\tilde r{_A^{4}}}}{G}H\left(\dot H-\frac{k}{a^2}\right).
\end{equation}
By using the Friedmann equations (\ref{frds21}) and (\ref{frds22}), we get
\begin{equation}
\dot S_A=3S_A H\,\frac{{\dot\phi^2}+3\omega^2H^2\dot\phi^2
+2\omega^2 k \dot\phi^2/a^2-\omega^2\frac{{\rm d}}{{\rm d}t}(H\dot\phi^2)+\rho_b+p_b}{\dot\phi^2(1+9\omega^2 H^2+3\omega^2 k/a^2)/2+V(\phi)+\rho_b}
=3S_A H \frac{\rho_{\rm {tot}}+p_{\rm {tot}}}{\rho_{\rm {tot}}}.
\end{equation}
As long as $\rho_{\rm {tot}}+p_{\rm {tot}}\ge 0$, we have $\dot S_A\ge 0$, and the second law of thermodynamics on the apparent horizon is satisfied.

\section{conclusions}

With the usual definition of the area law of entropy $S_A=\pi \tilde r_A^2/(4G)$ of the apparent horizon, and
the temperature $T_A=-[1-\dot{\tilde r}_A/(2H\tilde r_A)]/(2\pi\tilde r_A)$, as well as the energy conservation
for the effective total energy density $\dot\rho_{\rm {tot}}+3H(\rho_{\rm {tot}}+p_{\rm {tot}})=0$, we show that the
unified first law of thermodynamics ${\rm d}E=T_A{\rm d}S_A+W{\rm d}V$ is equivalent to the Friedmann equation for
the scalar--tensor theory with non-minimally derivative coupling. The result further supports the argument that the apparent
horizon is a physical boundary and the relation between the first law of thermodynamics and Friedmann equation
holds for more general theory of gravity and suggests a deep and fundamental connection between gravitation,
thermodynamics, and quantum theory.
Furthermore, we show that the second law of thermodynamics on the apparent horizon is also satisfied
for the scalar--tensor theory with non-minimally derivative coupling as long as $\rho_{\rm {tot}}+p_{\rm {tot}}\ge 0$.

\begin{acknowledgements}
This research was supported in part by the Natural Science
Foundation of China under Grant Nos. 11175270 and 11475065,
and the Program for New Century Excellent Talents in University under Grant No. NCET-12-0205.
\end{acknowledgements}


\end{document}